\documentstyle[referee]{l-aa}
\input{epsf}
\input{rotate}
\begin{document}

\thesaurus{04(02.07.1; 11.11.1; 03.13.1; 12.04.1)}
\title{An approximate
analytical method for inferring the law of gravity from the
macroscopic dynamics: 
Thin-disk mass distribution with exponential density}
\author{C. Rodrigo-Blanco}
\institute{Laboratorio de Astrof\'{\i}sica Espacial y F\'{\i}sica 
Fundamental (LAEFF),
P.O. Box 50727, E-28080 Madrid, Spain.\\
e-mail: crb@laeff.esa.es}
\date{Received date; accepted date}

\maketitle
\markboth{C.Rodrigo-Blanco: Non-Newtonian gravity directly from 
observations: thin-disk mass distribution.}{}

\begin{abstract}
The gravitational potential and the
gravitational rotation field generated by an
 thin-disk mass distribution with exponential density
are considered in the case when the force
 between any two mass elements is not the usual Newtonian one, but
 some general central force.
We use an approximation such that in the Newtonian case the
gravitational field generated by the disk reduces to the familiar
expression that results from applying the Gauss' law.
In this approximation, we invert the usual integral relations in such
 a way that the elemental interaction (between two point-like masses) 
is obtained as
a function of the overall gravitational field (the one generated
 by the distribution).
Thus, we have a direct way for testing whether it is possible 
or not to find a correction to the Newtonian law of gravity that 
can explain the observed dynamics in spiral galaxies without
dark matter.

\end{abstract}

\section{Introduction.}
In a previous paper (Rodrigo-Blanco \cite{Rodrigo96} :
hereafter, paper I )
we found the exact solution to the problem of,
given the gravitational field generated by an spherical mass
distribution with exponential density, finding out which 
was the elemental gravitational force, that is, the
force between any two point-like masses, that could have generated
that field. 

In this paper we follow the same steps as
in paper I , but the mass distribution that we consider is a thin
disk with exponential density, which is a more realistic description
of a spiral galaxy.

In this case, the exact solution cannot be found and 
the problem is solved in an approximation that, in the Newtonian case,
 is equivalent to use the Gauss' law for calculating the gravitational
field. We show that this is a very good approximation
which improves even more when the $g_{\rm eff}(r)$, that represents 
the 
deviation from the Newtonian force, is a growing function of $r$, 
which is a 
welcome behaviour for explaining the observed data. We also see
that this approximation is clearly better than using the results
obtained in paper I for the case of spherical symmetry.

In Section \ref{sec:general} we 
give the general definitions that are used later in Section 
\ref{sec:thin-disk}
for the case of a thin disk with exponential density.
In Section \ref{sec:development} we
show the mathematical basis underlying the results presented in 
Section \ref{sec:thin-disk}. Finally we offer some conclusions. In 
Appendix \ref{app:diskvsesf} we 
study the validity of the approximation. 
In Appendix \ref{app:math} we list some of the mathematical
identities used in Section \ref{sec:development}.



\section{General definitions}
\label{sec:general}
Let us assume that the gravitational
 potential ge\-ne\-ra\-ted by a point-like mass
does not correspond to the usual Newtonian form
but can be written in terms of a function $g(r)$ that 
describes the deviation from the Newtonian law, that is,

  \begin{equation}
	\phi (r) \equiv - \frac{G_0 m}{r}g(r).
  \label{punctpot}
  \end{equation}

The force per unit mass is, by definition, the gradient of
 the potential,

  \begin{equation}
	\vec{F}(r) \equiv - \frac{G_0 m}{r^2}g_{\rm eff}(r) 
\frac{\vec{r}}{r},
  \label{punctforce}
  \end{equation}
where we have introduced

  \begin{equation}
	g_{\rm eff}(r) \equiv g(r) - r g'(r).
  \label{ggef}
  \end{equation}

In this way, to find the total potential or the total force
generated by a mass distribution $\Omega$ with density
$\rho(\vec{r})$, it is necessary integrate over the volume 
spanned by $\Omega$ to get:

  \begin{equation}
	\Phi(\vec{R}) = - G_0 \int \int \int_{\Omega} d^3\vec{r} \  \ 
	\frac{g(|\vec{R}-\vec{r}|)}{|\vec{R}-\vec{r}|} \ \  
\rho(\vec{r})
  \label{potomeg}
  \end{equation}
for the potential experienced by a point mass at a distance $R$ 
from the centre of
$\Omega$, and
  \begin{equation}
	\vec{F}(\vec{R}) = - G_0 \int \int \int_{\Omega} 
d^3\vec{r} \ \  
	\frac{g_{\rm eff}(|\vec{R}-\vec{r}|)}{|\vec{R}-\vec{r}|^2}
 \ \  
	\frac{\vec{R}-\vec{r}}{|\vec{R}-\vec{r}|} \ \  \rho(\vec{r})
  \label{forceomeg}
  \end{equation}
for the force.

In the case that the gravitational potential is only a function of the
distance to the centre of the distribution,
it is convenient to introduce two new functions $\Psi(R)$ 
and $\Psi_{\rm eff}(R)$ such that:

  \begin{equation}
	\Phi(R) \ \equiv \ - \ \frac{G_0 M_{\rm tot}}{R} \ \Psi(R),
  \label{potpsi}
  \end{equation}

  \begin{equation}
	\vec{F}(R) \ \equiv \ -\frac{G_0 M_{\rm tot}}{R^2},
	 \ \Psi_{\rm eff}(R) \ \frac{\vec{R}}{R}
  \label{v2psi}
  \end{equation}
and the rotation velocity of a test particle in a circular orbit bound
to the distribution will be:
\begin{equation}
	V^2_{\rm rot}(R) \ \equiv \ \frac{G_0 M_{\rm tot}}{R}
	 \ \Psi_{\rm eff}(R)
  \label{v2psi2}
  \end{equation}
where the auxiliary functions $\Psi_{\rm eff}(R)$ and $\Psi(R)$ 
satisfy 
the following relationship:

  \begin{equation}
	\Psi_{\rm eff}(R) = \Psi(R) -R \Psi'(R),
  \label{psiphi}
  \end{equation}

Our goal is to design a procedure where, assuming that 
$\vec{F}(R)$ is known
 (say from observation of the rotation velocity) for all values of $R$,
a $g_{\rm eff}(r)$ that generates the given rotation 
velocity is obtained. 
In other words, given the potential as inferred from observations 
we want to find which $g(r)$ 
could have generated it. Actually, we find $g(r)$ and 
$g_{\rm eff}(r)$ as
functions of $\Psi_{\rm eff}(R)$ and $\Psi(R)$ respectively.

\section{Thin-disk mass distribution with exponential density}
\label{sec:thin-disk}

The luminosity
profile of many spiral galaxies can be well fitted assuming that the  
luminous matter is placed along a thin disk with a density 
that decreases exponentially with the distance to the
centre of the galaxy (Kent \cite{Kent87}).

\begin{equation}
	\rho(r) \equiv \rho_0 \ e^{-\alpha r}.
\label{density}
\end{equation}

Considering a thin-disk distribution and using Eqs. 
(\ref{potpsi}), (\ref{v2psi})  and (\ref{density}), 
in Eqs. (\ref{potomeg}) and 
(\ref{forceomeg}), the two problems outlined above 
can be recast as two integral equations:

\vspace{4mm}

\noindent(i)	Given  $\Psi(R)$, defined by (\ref{potpsi}), 
find a function 
	$g(r)$ such that:

	\begin{equation}
		\int_0^{\infty}dr\int_0^{\pi}d\theta\int_0^{2\pi}d\phi
		\frac{g(\sqrt{R^2+r^2-2Rr\cos\theta})}
		{\sqrt{R^2+r^2-2Rr\cos\theta}}
		r^2 \sin\theta e^{-\alpha r} = \frac{8 \pi}{\alpha^3}
		\frac{\Psi(R)}{R}. \\
	\label{probpotdisk}
	\end{equation}

\vspace{4mm}

\noindent(ii)	Given $\Psi_{\rm eff}(R)$, defined by 
(\ref{v2psi}), find a function
	 $g_{\rm eff}(r)$   such that:

	\begin{equation}
		\int_0^{\infty}dr\int_0^{\pi}d\theta
\int_0^{2\pi}d\phi\frac{g_{\rm eff}
		(\sqrt{R^2+r^2-2Rr\cos\theta})}
{(R^2+r^2-2Rr\cos\theta)^{\frac{3}{2}}}
		(R-r\cos\theta) r^2 \sin\theta e^{-\alpha r} = 
		\frac{8 \pi}{\alpha^3}\frac{\Psi_{\rm eff}(R)}{R^2}.
	\label{probforcedisk}
	\end{equation}

In this case  the problem cannot be solved exactly. We use an
approximation that we call  {\it Gaussian} as, in the Newtonian case,
it is equivalent to use the Gauss' law for calculating 
the gravitational field.

In the next section the calculations are described in detail. The 
solution to the two problems outlined above, in the {\it Gaussian
approximation} can be summarized as:

\noindent  (i) Potential problem:
({\it viz.} Eqs. (\ref{potpsi}) and (\ref{probpotdisk}))

	In this case, the approximate solution to the problem is 

	\begin{equation}
		g(x)=  \Psi(x)-\frac{1}{\alpha^2}\Psi''(x),
	\label{gpsi}
	\end{equation}
       where the function $\Psi$ has the following 
behaviour at the origin:

	\begin{equation}
	\Psi(0) \ =  \ 0.
	\label{gpsicond}
	\end{equation}

\noindent(ii)	Force and velocity problem:
({\it viz.} Eqs. (\ref{v2psi}), (\ref{v2psi2}) and 
	(\ref{probforcedisk})).

         Here, the approximate solution is given by the 
following expression:

	\begin{equation}
		g_{\rm eff}(x)= 
		 \Psi_{\rm eff}(x)-\frac{1}{\alpha^2}
		\Psi_{\rm eff}''(x) 
		+ \frac{2}{\alpha^2 x}\Psi_{\rm eff}'(x)  
	\label{geffpsieff}
	\end{equation}
and the behaviour of $\Psi$ at the origin is as follows:

	\begin{equation}
		\Psi_{\rm eff}(0) \ = \ \Psi_{\rm eff}'(0) \ = \ 0.
	\label{geffpsieffcond}
	\end{equation}


\section{Mathematical formalism.}
\label{sec:development}

First we show how to solve {\it the potential
problem}, that is, how to go from equation (\ref{probpotdisk})
 to equations (\ref{gpsi}) and (\ref{gpsicond}). Then, the results 
obtained above are used to tackle {\it the
force problem}, i.e, to go from
equation (\ref{probforcedisk}) to equations (\ref{geffpsieff})
 and (\ref{geffpsieffcond}).

\subsection{The potential.}
\label{subsec:potential}

In order to go from equation Eq. (\ref{probpotdisk}) to Eqs.
 (\ref{gpsi}) and (\ref{gpsicond}), it is necessary to use several
 mathematical identities. For convenience, these are listed in 
Appendix \ref{app:math}.

Using (\ref{fourier}), (\ref{addth}),(\ref{addth2}) and
 (\ref{Legendre})
 into (\ref{probpotdisk})  $\Psi(R)$ can be written as:

\begin{equation}
\begin{array}{rl}
	\Psi(R)= & - \frac{\alpha^2 }{\pi \sqrt{R} } 
	\sum_{\rm k=0}^{\infty} (4k+1)
	\left[\frac{\Gamma(2k+1/2)}{\Gamma(2k+1)}\right]^2
	\int_0^{\infty}dp \frac{\hat{g}_s(p)}{p} J_{\rm 2k+1/2}(pR)
	 \int_0^{\infty} dr e^{-\alpha r} \sqrt{r}
	J_{\rm 2k+1/2}(pr) \\
\end{array}
\label{sumtot}
\end{equation}

We apply our approximation here. We only consider
the first term in the series of Bessel functions, that is, we 
drop all the terms in the series but the one with $k=0$. It can be
seen, and we prove it in an appendix, that in the Newtonian limit
(i.e, when $g(r)$ = $g_{\rm eff}(r)$ = $1$) this 
approximation corresponds 
to apply the Gauss' law to the mass distribution, i.e, to say that the
gravitational force at a distance $R$ is proportional to the total
mass contained in the sphere of radius $R$. This is the reason
why we have called this method {\it Gaussian
approximation}. As it is quite a good approximation in the Newtonian
case, and we also will see that it is exact for some forms of  $g(r)$
 and a very good approximation for the most interesting 
forms of $g(r)$,
 we restrict ourselves to consider only the first term in the series.

Thus, we will write:

\begin{equation}
	\Psi_0(R)=  - \frac{\alpha^2 }{\sqrt{R} } 
	\int_0^{\infty}dp \frac{\hat{g}_s(p)}{p} J_{\rm 1/2}(pR)
	 \int_0^{\infty} dr e^{-\alpha r} \sqrt{r}
	J_{\rm 1/2}(pr)
\label{Psi0}
\end{equation}

Now, using the functional form of $J_{\rm 1/2}$ and applying
 (\ref{fourierinv}) to invert the Fourier transformation, and after
some straightforward calculations, we get:
\begin{equation}
	\Psi_0(R)= - \frac{\alpha}{2} 
		\left\{ 
		e^{\alpha R} \left[ \int_0^R dx g(x) e^{-\alpha x} -
				\int_0^{\infty} dx g(x) e^{-\alpha x}
			     \right] 
		- e^{-\alpha R}\left[ \int_0^R dx g(x) e^{\alpha x} -
				\int_0^{\infty} dx g(x) e^{-\alpha x}
			     \right] \right\}.
\label{potexpon}
\end{equation}

At this point, it is useful to introduce 
an au\-xi\-lia\-ry function 
$\psi(x)$ that makes the integrals exact:

\begin{equation}
	g(x) \ \equiv \ \psi(x) - \frac{1}{\alpha^2} \psi''(x). 
\label{gpsieq}
\end{equation}

Eq (\ref{gpsieq}) can be inserted into Eq. (\ref{potexpon}) and, 
upon integration by parts, we get:

\begin{equation}
	\Psi_0(R) = \psi(R) - e^{-\alpha R} \psi(0),
\label{Psipsi}
\end{equation}
where $\psi$ is a solution to the ordinary differential Eq.
 (\ref{gpsieq}) that satisfies the conditions
 of being
 an analytic function at $x=0$, and

\begin{equation}
	\lim_{x \rightarrow \infty}\psi(x) \exp(-\alpha x) = 
	\lim_{x \rightarrow \infty}\psi'(x) \exp(-\alpha x) = 0.
\label{infcond}
\end{equation}

These conditions are easily satisfied in all the astrophysical systems.
The analiticity is satisfied in the Newtonian limit, which is
the behaviour that we expect to recover at $R \sim 0$ and it can be
seen that if Eq. (\ref{infcond}) were not satisfied, the rotation
velocity would grow almost exponentially with the distance, which
clearly seems to contradict the observations.

It is straightforward to see that, provided 
$\psi(R)$ is a solution to Eq. (\ref{gpsieq}), 
then $\Psi_0(R)$ is also a
solution to the same equation. Moreover, the term proportional to
$\psi(0)$ in Eq. (\ref{Psipsi}) implies that $\Psi_0$ is
 zero at the origin. Taking all that into
account, we finally obtain:

\begin{equation}
	 g(x) = \Psi(x)-\frac{1}{\alpha^2}\Psi''(x),
\label{finalPhi}
\end{equation}
and
\begin{equation}
	\Psi(0) \ = \  0.
\label{finalPhi2}
\end{equation}
where we have omitted the subscript $0$ everywhere as 
we will in what follows.


\subsection{The force and the rotation velocity.}
\label{subsec:velocity}

Once $\Psi(R)$ is known, $\Psi_{\rm eff}(R)$ 
can be calculated using Eq. (\ref{psiphi}).
Equivalently, once $g(r)$ is known, $g_{\rm eff}(r)$ can 
be obtained through Eq.
(\ref{ggef}). Using these two equations together with Eq. (\ref{finalPhi}), 
and after some straightforward calculations, a direct
relation between $g_{\rm eff}$ and $\Psi_{\rm eff}$ can be obtained:

\begin{equation}
	g_{\rm eff}(x)=  \Psi_{\rm eff}(x)-\frac{1}{\alpha^2}
\Psi_{\rm eff}''(x)
		 + \frac{2}{\alpha^2 x}\Psi_{\rm eff}'(x).
\label{geffpsieff2}
\end{equation}

From Eq. (\ref{psiphi}) and the behaviour of $\Psi$ at the origin, Eq.
(\ref{finalPhi2}), it is easy to see
that, at the origin, $\Psi_{\rm eff}$ will satisfy:

\begin{equation}
      \Psi_{\rm eff}(0) \ = \ \Psi'_{\rm eff}(0) \ =  \ 0.
\label{psieffcond}
\end{equation}

\section{Summary and conclusions.}

We have found the solution to the problem of inverting the integral
 relation between the elemental law of gravity and the overall 
gravitational
field generated by a thin-disk mass distribution. We have done it in
an approximation that we have called {\it Gaussian} as it is
equivalent to use the Gauss' law for calculating the gravitational
field generated by the distribution. Although this is not exact, we
show in an appendix that it is a very good approximation, and it gets
 much better when
$g_{\rm eff}(r)$ grows with $r$, which is the expected behaviour if 
the observed gravitational behaviour must be explained without the 
need of dark matter.

In summary, we now have a method for inferring $g_{\rm eff}(r)$ 
given $\Psi_{\rm eff}(R)$.
It can be said that we have a {\it way to travel from the world of
macroscopic interactions to the world of microscopic interactions}.
It can be schematized as follows:
Given the observed $V_{\rm rot}(R)$ for a given galaxy, use
Eq. (\ref{v2psi2})
to obtain $\Psi_{\rm eff}(R)$, fit it by a mathematical 
function and then
use Eq. (\ref{geffpsieff}) to get the $g_{\rm eff}(r)$ 
that describes the
elemental gravitational force (through Eq. 
(\ref{punctforce})) that can
explain the observed rotation curve.

We are now ready for applying the differential expression that we have
found to the observed rotation curves of spiral galaxies, and find the
$g_{\rm eff}$ required for explaining those rotation curves. Doing
that, we will see whether it is possible or not to 
find a universal law of
gravity that can explain all the rotation  curves  
needing only
the observed luminous matter. This will be done in a separate publication 
(Rodrigo-Blanco \& P\'erez-Mercader \cite{RodrigoPerez96}).

\appendix
\section{The validity of the approximation.}
\label{app:diskvsesf}

In the step from Eq. (\ref{sumtot}) to Eq. (\ref{Psi0}) we  made the
approximation of considering only the first term of an infinite series.
The terms dropped depend also on $g(r)$, which is not known,
 so  it is not trivial to evaluate the goodness of the approximation.

In this appendix, we  choose some functions $g_{\rm eff}(r)$ 
and  compare
the {\it exact} rotation velocity with the {\it approximate} 
one. By {\it
exact} we mean the rotation velocity generated by the considered 
$g_{\rm eff}(r)$
through Eqs. (\ref{probforcedisk}) and (\ref{v2psi2}). 
Nevertheless, for most functional forms of
$g_{\rm eff}(r)$ an analytical solution cannot be found and 
it is necessary
 to perform numerical integration for finding that {\it exact} 
rotation velocity. Whenever that happens we have chosen to perform the
integrals assuming that 
the disk has some non-zero thickness. Actually this is a
 more realistic model
for a spiral galaxy, the astrophysical system to which our
 approximate method
is applied. By {\it approximate} rotation velocity we
mean the one such that the corresponding $\Psi_{\rm eff}(R)$ satisfies
Eqs. (\ref{geffpsieff2}) and (\ref{psieffcond}).

As a test of consistency we first consider the Newtonian case, i.e.,
$g(r)=g_{\rm eff}(r)=1$. When we use
Eqs. (\ref{geffpsieff2}) and (\ref{psieffcond}) we obtain:

\begin{equation}
	\Psi_{\rm eff}(R) = 1-(1+\alpha R) e^{-\alpha R},
\label{psieffg=1}
\end{equation}
and thus, the rotation velocity is
\begin{equation}
	V^2_{\rm rot}(R) = \frac{G_0M_{\rm tot}}{R}
			[1-(1+\alpha R) e^{-\alpha R}]
		     = \frac{G_0M(R)}{R},
\label{v2g=1}
\end{equation}
where $M(R)$ is the disk mass inside the sphere of radius $R$. Of
course, this is not the exact result, but it is what we find if
we apply the Gauss' law as an approximation for evaluating the
gravitational field. That is why the 
approximation is called {\it Gaussian}.

Next, our approximation must be checked for other 
different forms of $g_{\rm eff}$.
For doing that we choose a parametric family of 
$g_{\rm eff}(r)$'s given by

\begin{equation}
g_{\rm eff,\mu}(r) \equiv \left(\frac{r}{a}\right)^{\mu},
\label{gmu}
\end{equation}
where $\mu$ parametrizes how fast $g_{\rm eff}$ grows.

\begin{figure*}
\epsfxsize=10cm
\epsfbox{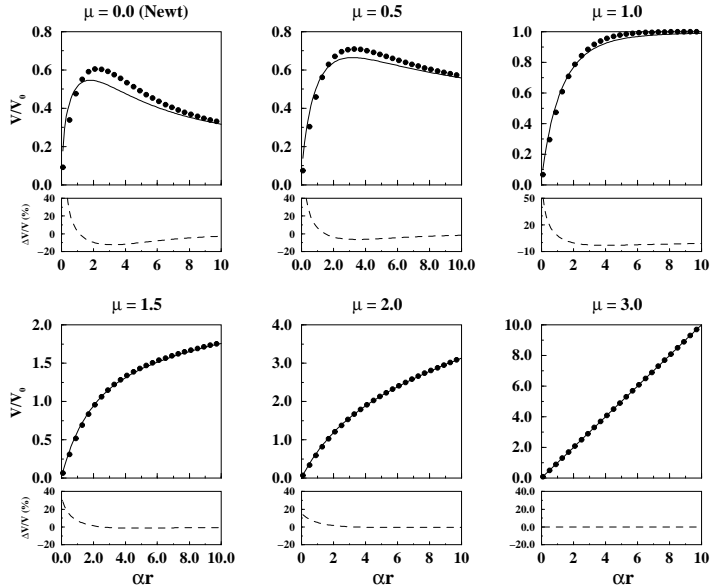}

\caption{Exact rotation curves (dots) and {\it Gaussian}
approximation (solid line) when $g_{\rm eff}(r) \equiv
\left( \frac{r}{a}\right)^{\mu}$ for
some values of $\mu$ ($\mu = 0$ is the Newtonian case). In every
case, for the sake of clarity, the velocities are 
normalized by dividing
by an appropriate constant 
${\rm V}_0 \equiv \frac{G_0 M_{\rm tot}\alpha}{(\alpha a)^{\mu}}$.
In the inset
graphs (dashed lines) we have plotted, as a function 
of $r$ for each case,
 the percentage of
 error made when the Gaussian approximation is used instead
of the numerical integrals.
\label{fig:comparison}}
\end{figure*}

Concerning the {\it exact} solution we have found the analytical one
(for the thin-disk case) for $\mu=1$ and $\mu=3$:
\begin{equation}
	V^2_{\rm rot,\mu=1}(R)= \frac{G_0 M_{\rm tot}}{a}
[1-(1+\alpha R)e^{-\alpha R}]
			  = \frac{G_0 M(R)}{R} \ \frac{R}{a},
\label{Vmu1}
\end{equation}
\begin{equation}
	V^2_{\rm rot,\mu=3}(R)= \frac{G_0 M_{\rm tot}}{R} \
			\left(\frac{R}{a}\right)^3 \left( 1 +
			\frac{6}{\alpha^2 R^2} \right).
\label{Vmu3}
\end{equation}

For the other values of $\mu$ the integrals are performed numerically
for a disk with a small thickness $h=\alpha^{-1}/6$

Concerning the approximation, it can be seen that, for $\mu=3$ the
solution is (\ref{Vmu3}), that is, the approximation is exact. 

 In Figure \ref{fig:comparison}
we have plotted the exact rotation velocities compared with those
corresponding to our approximation for some values of $\mu$. 
It can be seen
that the results are very similar in every case.
For each case, the solutions are normalized
dividing by  a convenient constant $V_0$ defined as:

\begin{equation}
V_0 \equiv \frac{G_0M_{\rm tot}\alpha}{(\alpha a)^{\mu}}.
\label{defV0}
\end{equation}

\begin{figure}
\setbox0=\hbox{
\epsfxsize=6.5cm
\epsfbox{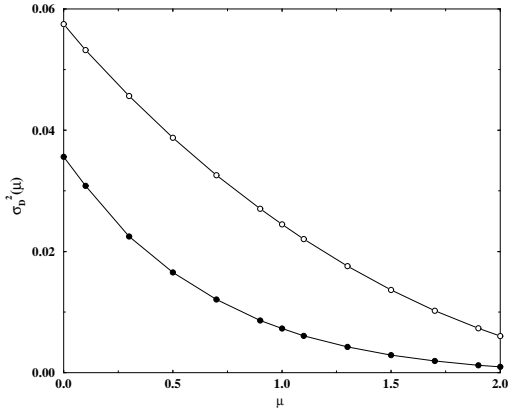}
}
\rotr 0

\caption{Mean square error in the velocity 
when the {\it Gaussian} approximation is used instead of the exact
solution for the disk, as a function of $\mu$, 
 and using $g_{\rm eff}(r) \equiv \left(\frac{r}{a}\right)^{\mu}$. 
The exact
meaning of $\sigma^2_{\rm D}$ is explained in the text.}
\label{sigmaDgauss}
\end{figure}

Although it can be seen in Figure \ref{fig:comparison} 
that the approximation is quite good in every
case, and is better when $g_{\rm eff}(r)$ is a growing function
 of $r$, it is
interesting to have a more quantitative way of describing the 
difference in the
rotation velocity obtained in the exact case and in the {\it Gaussian}
approximation as a function of $\mu$. In order to do that, we define a 
quantity $\sigma^2_D$ as follows:

\begin{equation}
	\sigma^2_D (\mu) \equiv \frac{1}{N} \sum_{\rm i=1}^{N} 
	\frac{(V_{\rm D,\mu}(r_i)-V_{\rm G,\mu}(r_i))^2}
	{V^2_{\rm D,\mu}(r_i)},
\label{sigmaDdef}
\end{equation}
where the sub-scripts $D$ and $G$ stand for the disk (exact) and the
{\it Gaussian} approximation,
res\-pec\-ti\-ve\-ly. We sum over $r_i$, which are the points where the
integrals are calculated. The total number of points for each 
value of $\mu$ is $N=100$.

So defined, $\sigma^2_D(\mu)$ is a measure of the mean square
 error that we  make in the rotation velocity if the approximation 
is used instead of the numerical integrals, for each value of $\mu$.

In figure (\ref{sigmaDgauss}), we plot the value
of $\sigma^2_D$ versus $\mu$. Once again, it can be seen that the
larger $\mu$ is, the smaller the difference.
We also plot the values of $\sigma^2_D$ obtained in paper I
when the equations for spherical symmetry were used as an approximation
to the disk problem. It can be seen that the {\it Gaussian}
approximation is always better than the spherical one.


\section{Appendix: Mathematical identities}
\label{app:math}

  \begin{enumerate}

  \item{ Fourier sine transform}
	\begin{equation}
	 g(\sqrt{r^2+R^2-2rR\cos{\theta}}) \equiv  \frac{2}{\pi}
	  \int_0^{\infty}\hat{g}_s(p) \sin(p\sqrt{r^2+R^2-2rR\cos
	  {\theta}}) dp \label{fourier}
	\end{equation}

	\begin{equation}
	  \hat{g}_s(p) = \int_0^{\infty} g(t) \sin{pt} dt 
	\label{fourierinv}
	\end{equation}

  \item{ Addition theorem for Bessel functions} (see Gradshteyn
\cite{Gradshteyn80}).

	\begin{equation}
	    	\frac{Z_{\nu}(m\omega)}{\omega^{\nu}}= 
	\frac{2^{\nu}}
    	{m^{\nu}}\Gamma(\nu)\sum_{\rm k=0}^{\infty}(\nu +k)
	       \frac{J_{\rm \nu+k}(m\rho)}{\rho^{\nu}} 
\frac{J_{\rm \nu+k}(mr)}
	   	{r^{\nu}}C_k^{\nu}(\cos{\theta})  
	\label{addth}
	\end{equation}
	where:
	\begin{equation}
	   	  \left\{\begin{array}{l}
		\omega \equiv \sqrt{r^2+\rho^2-2r\rho\cos{\theta}} \\
		\rho < r \\
		C_k^{\nu} \equiv \mbox{Gegenbauer Polynomials.}
		\end{array}\right. 
	\label{definitions}
	\end{equation}

\vspace{4mm}

\noindent	Using Eq. (\ref{addth}), together with:
	\begin{equation}
	\sin (mz) = \sqrt{\frac{\pi m z}{2}} J_{\rm 1/2}(mz)
	\label{sinbessel}
	\end{equation}

\vspace{4mm}

\noindent	we get:

	\begin{equation}
	\frac{\sin{p\sqrt{r^2+R^2-2rR\cos{\theta}})}}
	 {\sqrt{r^2+R^2-2rR\cos{\theta}}} = 
	\frac{\pi}{2\sqrt{Rr}}\sum_{\rm k=0}^{\infty}
	 (2k+1) J_{\rm k+\frac{1}{2}}(pr) 
	J_{\rm k+\frac{1}{2}}(pR) P_k(\cos{\theta})
	\label{addth2}
	\end{equation}

  \item{ Legendre Polynomials.}
	\begin{equation}
	\int_0^{2\pi} d\theta P_k(\cos{\theta})  =
	\left\{ \begin{array}{lcl}
	         2\left[\frac{\Gamma(2k+1/2)}{\Gamma(2k+1)}\right]^2
			& ; & \mbox{if } k=2n+1 \\
		&& \\
		0  & ; & \mbox{if } k=2n \\
		\end{array} \right.
	\label{Legendre} 
	\end{equation}

  \end{enumerate}

\acknowledgements{I am grateful to Juan P\'erez-Mercader
for his guidance and help, and also to Massimo Persic 
and Paolo Salucci for their hospitality at SISSA, where a 
part of this work has been done, and for their helpful comments.}


\end{document}